\documentclass{optica-article}

\journal{opticajournal} 

\articletype{Research Article}

\begin{document}

\title{Monolithic silicon nitride electro-optic modulator enabled by optically-assisted poling}

\author{Christian Lafforgue,\authormark{1} Boris Zabelich,\authormark{1, 2} and Camille-Sophie Brès\authormark{1,*}}

\address{\authormark{1}Ecole Polytechnique Fédérale de Lausanne, Photonic Systems Laboratory (PHOSL), CH-1015 Lausanne, Switzerland\\
\authormark{2}Now at: LIGENTEC SA, EPFL Innovation Park, CH-1024 Ecublens, Switzerland}

\email{\authormark{*}camille.bres@epfl.ch} 

\begin{abstract*}
Electro-optic (EO) modulation is a key functionality to have on-chip. However, achieving a notable linear EO effect in stoichiometric silicon nitride has been a persistent challenge due to the material's intrinsic properties. Recent advancements revealed that the displacement of thermally excited charge carriers under a high electric field induces a second-order nonlinearity in silicon nitride, thus enabling the linear EO effect in this platform regardless of the material's inversion symmetry. In this work, we show for the first time optically-assisted poling of a silicon nitride microring resonator, removing the need for high-temperature processing of the device. The optical stimulation of charges avoids the technical constraints due to elevated temperature. By optimizing the poling process, we experimentally obtain a long-term effective second-order nonlinearity $\chi^{(2)}_{\rm{eff}}$~of 1.2~pm/V. Additionally, we measure the high-speed EO response of the modulator, showing a bandwidth of 4~GHz, only limited by the quality factor of the microring resonator. This work goes towards the implementation of monolithic, compact silicon nitride EO modulators, a necessary component for high-density integrated optical signal processing.
\end{abstract*}

\section{Introduction}

Stoichiometric silicon nitride (Si$_3$N$_4$) is massively studied and developed by researchers and industries due to its important potential for large-scale production of photonic integrated circuits (PIC). Linear functionalities like wavelength multiplexing and demultiplexing, filtering, pulse manipulation or polarization management are now commonly established in industrial-grade manufactures using this platform \cite{Xiang_2022, Wilmart_2019, Blumenthal_2018, Rahim_2017}. Propagation losses of Si$_3$N$_4$~waveguides currently reach values as low as a few~dB/m \cite{Liu_2021, ElDirani_2019} and coupling losses are reduced to the~dB~level~per~facet, allowing extending applications of the platform to frequency conversion, soliton microcombs, spectral shaping and optical switching exploiting nonlinear optical effects. Researchers recently demonstrated on-chip amplification and lasing in Si$_3$N$_4$~PICs \cite{Liu_Qiu_2022, Liu_Qiu_Ji_2024}, solving one of the major bottlenecks of this platform. One of the last remaining building blocks for optical signal processing in Si$_3$N$_4$~PICs is the electro-optic (EO) modulator. Due to the high bandgap of amorphous Si$_3$N$_4$~and its inherent inversion symmetry, commonly used methods based on plasma dispersion effect (as in silicon modulators) or Pockels effect (as in lithium niobate modulators) cannot be implemented in this platform. 
Nevertheless, several published studies in the last few years demonstrated the ability to induce a second-order nonlinearity ($\chi^{(2)}$) thanks to charge relocation in a waveguide's core, first resulting in second harmonic generation by all-optical poling \cite{Yakar_2022, Nitiss_2020, Hickstein_2019} and then in EO modulation by thermally-assisted poling \cite{Zabelich_Nitiss_2022, Zhang_Nauriyal_2023, Zabelich_Lafforgue_2024}. The latter approach involves high voltage application through patterned electrodes at elevated temperature to inscribe a built-in constant electric field in the waveguide's core. This field coupled with the third-order nonlinearity ($\chi^{(3)}$) then creates an effective second-order nonlinearity ($\chi^{(2)}_{\rm{eff}}$). 
However, the highest reported values of $\chi^{(2)}_{\rm{eff}}$~have remained in the sub-pm/V level so far, and the necessity of exposing the devices to high temperature brings complexity to the poling process. Indeed, such a process usually requires a specific setup and reduces the dielectric strength of the materials and the air surrounding the contact pads \cite{Sze_1967, Klein_Gafni_1966}, thereby limiting the maximum voltage that can be applied before dielectric breakdown occurs. Inspired by the works on electrically poled fibers assisted by intense light excitation \cite{Bergot_1988, Fleming_An_2008, Corbari_2005, Camara_2015, Pereira_2019}, we implemented a new method for poling of Si$_3$N$_4$~waveguides. This technique utilizes visible light to excite charge carriers while applying high voltage to separate them within the waveguide's cross-section. By injecting light at 780~nm in a microring resonator, we achieved an effective second-order nonlinearity ($\chi^{(2)}_{\rm{eff}}$) of 1.2 pm/V, which is the highest reported for Si$_3$N$_4$~integrated waveguides to our knowledge. This result is currently limited by the voltage capacity of the existing electrode design. Upon optimization of the electrodes, a maximum value as high as 6.5~pm/V is expected. Moreover, the induced nonlinearity remains stable over time, with no significant degradation observed during 60 days. To assess the performance of this modulator, we conducted high-speed modulation tests, measuring an EO 3~dB bandwidth of up to 4 GHz, limited only by the photon lifetime of the ring resonator. This modulator addresses a key technological gap in Si$_3$N$_4$~photonic platforms for optical communications and signal processing.

\section{Principle of operation}

\subsection{Microring modulator}

In a ring resonator (RR), modulation is obtained through the resonance shift $\Delta f_r$ caused by the applied voltage ($U$)
\cite{Witzens_2018}:
\begin{equation}
    \Delta f_r(U) = f_r(0)-f_r(U)=\dfrac{f_r(0)\Delta n_{\rm{eff}}(U)}{n_g}
    \label{eq:delta_f},
\end{equation}
where $f_r(U)$ is the voltage dependent resonance frequency and $n_g$ is the group index of the ring waveguide. $\Delta n_{\rm{eff}}$, the change in effective index, is impacted by the local refractive index change, $\Delta n$, directly related to the nonlinear properties of the material \cite{Kogelnik_1988, Steglich_2020}:

\begin{equation}
\begin{split}
    &\Delta n_{\rm{eff}}(U)=\dfrac{n_0}{n_{\rm{eff}}}\dfrac{\iint_{\rm{core}}\Delta n(U)\cdot |E_{\rm{opt}}|^2dxdy}{\iint_{\infty}|E_{\rm{opt}}|^2dxdy},\\
    &\Delta n(U)=\dfrac{3\chi^{(3)}\left(E_{\rm{el}}(1~V)\cdot U\right)^2}{2n_0}+\dfrac{\chi^{(2)}E_{\rm{el}}(1~\rm{V})\cdot U}{n_0},
\end{split}
\label{eq:delta_n}
\end{equation}
where $E_{\rm{opt}}$ is the optical mode distribution, $E_{\rm{el}}(U)$ is the externally applied electric field for a voltage $U$ and $n_0$ is the initial material's refractive index. 

\begin{figure*}[t]
    \centering
    \includegraphics[width=\textwidth]{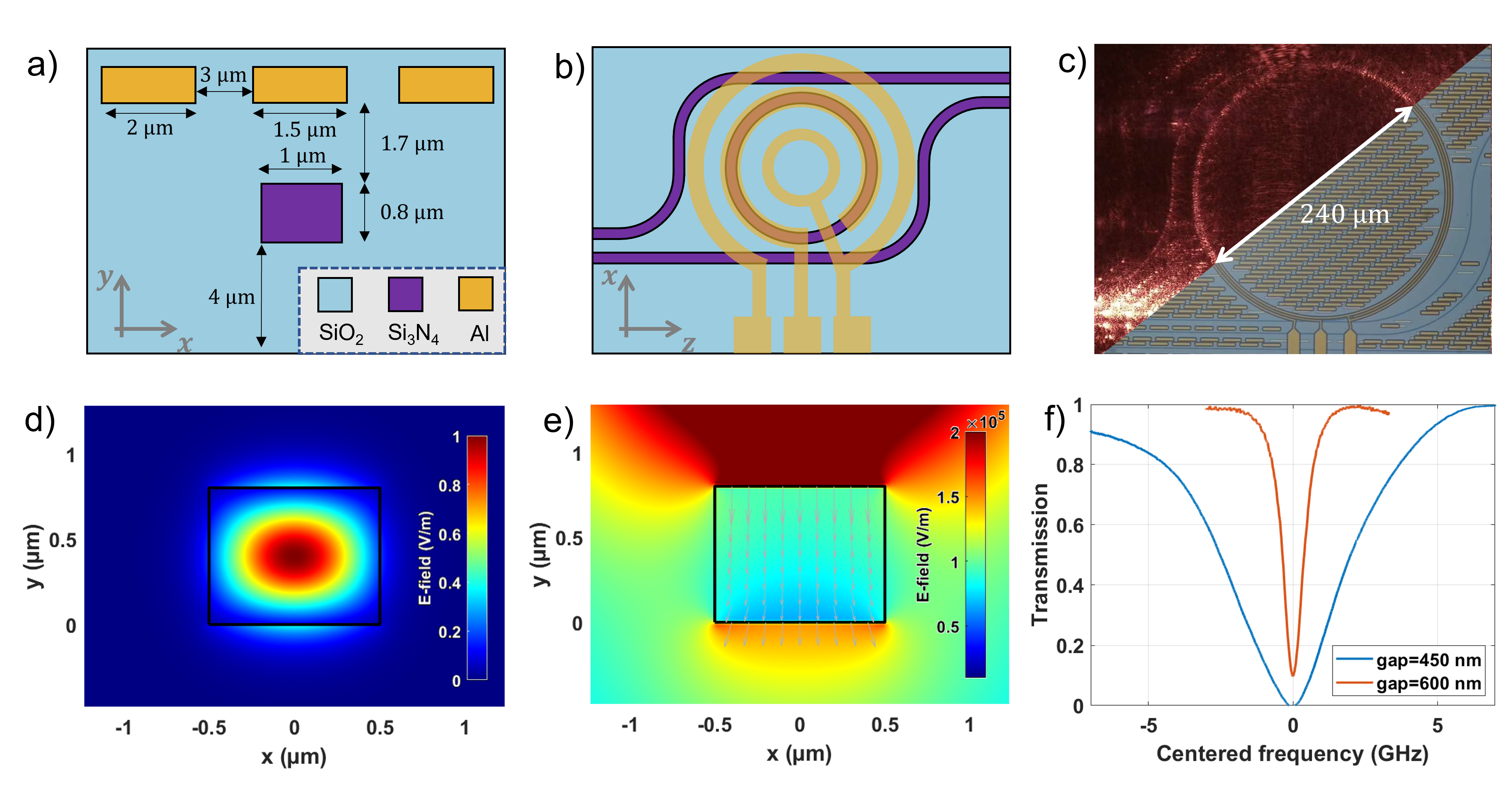}
    \caption{a) Schematics of the cross-section of the device. b) Top-view schematic of the microring modulator. c) Top view of the ring during optically assisted poling under a microscope, with the bottom right part showing the microscope illumination and the top left part showing the ring without illumination.  d) Optical mode distribution for the TM mode at a wavelength of 1550 nm. e) Static electric field distribution (vertical component) for 1 V applied. f) Example of resonances of the fundamental TM mode at a wavelength near 1550 nm for two different gaps between the bus and the ring. }
    \label{fig:fig1}
\end{figure*}

Fig.~\ref{fig:fig1} a) and b) show schematic views of the cross-section and top view of the device, respectively. To gain control and repeatability on the properties of the resonances, we opted for an add-drop RR configuration since, for low-loss rings, the Q-factor is primarily governed by the coupling strength between the buses and the ring. With a symmetric gap between the input waveguide and the drop waveguide, a nearly critical coupling is expected on all resonances for any gap regardless of loss variations, which allows to control the resonance linewidth without impacting the visibility. The chips were fabricated by LIGENTEC SA, providing low-loss ($\approx 0.2 $ dB/cm) Si$_3$N$_4$. The Si$_3$N$_4$~thickness is fixed to 800~nm by the LIGENTEC SA process, while we can select the waveguide's width to have a single-mode operation in the C-band window. A width of 1~\textmu m ensures that only fundamental TE and TM modes are supported by the waveguide while maintaining low propagation loss. The low bending loss provided by the 800~nm Si$_3$N$_4$~platform thanks to the high modal confinement allows to design RRs with a small radius to reduce the footprint of the device. We used rings with a diameter of 240 \textmu m, giving a free spectral range of 188 GHz. Coplanar metal electrodes were patterned in the cladding above the Si$_3$N$_4$ layer to apply a strong vertical electric field in the Si$_3$N$_4$~section. Because the nonlinear interaction is higher when the fields involved in the process are collinear \cite{Boyd}, the device was optimized to work in TM polarization. Fig.~\ref{fig:fig1} d) and e) give the fundamental TM optical mode and external electric field distributions respectively. The vector field in grey in Fig.~\ref{fig:fig1} e) shows that the external electric field is almost perfectly vertical across the waveguide's core and the average vertical field in the core for 1 V applied between the center and external electrodes is calculated to be 88.2 V/mm. The maximum electric field in the core aligns with the peak optical intensity, thereby optimizing the EO coupling. Low-power frequency sweeps were performed to characterize the resonances of the rings. Typical resonances in TM polarization for two different gaps between the bus and the ring are depicted in Fig.~\ref{fig:fig1} f). For this device, the Q-factor is $3.9\cdot 10^4$ and $2.5\cdot 10^5$ in TM polarization, for gaps of 450~nm and 600~nm respectively. The control on the Q factor allows us to adapt the device to the targeted application. While lower Q values are more suited for high speed EO modulation, the characterization of the resonance shift behavior in the DC regime is more precise with high Q values.

\subsection{Poling}

To endow SiN with the Pockels effect, and hence linear EO modulation, it has been shown that the material can be poled as to inscribe a persistent electric field $E_{\rm{ins}}$ within the SiN waveguides. The inscribed field coupled to the inherent third-order susceptibility of the material $\chi^{(3)}$ gives rise to an effective second-order susceptibility $\chi^{(2)}_{\rm{eff}}$$=3$$\chi^{(3)}$$E_{\rm{ins}}$. The main hypothesis behind the origin of this inscribed field is the displacement of trapped charge carriers, after activation, under an applied high external field. Initial work in integrated waveguides have achieved this activation through thermal excitation, either by applying high temperature to the sample or with a high power CO$_2$ laser. Alternatively, studies in fibers showed that the field inscription can also be obtained at room temperature by optical excitation using high-intensity light in the UV and visible ranges \cite{Bergot_1988, Fleming_An_2008, Corbari_2005, Camara_2015,Pereira_2019}. 
The setup used to pole our devices is shown in Fig.~\ref{fig:fig2} a). Light from a tunable continuous wave laser near 780 nm is coupled to the device and tuned to a resonance of the ring. The device is designed to operate at telecom wavelenths around 1550 nm, we leverage the multimode nature of the waveguide at 780 nm to couple the light in the ring. While consequently it is not possible to experimentally estimate the coupled power to the ring, we ensure that the maximum power is coupled by monitoring the sample through a microscope and observing the scattering as shown in Fig.~\ref{fig:fig1} c). The bottom right part is an image with the illumination of the microscope on, showing the ring and electrodes; the top left image is obtained with the microscope illumination off, clearly revealing the 780~nm optical stimulation propagating in the ring. With the 780 nm circulating in the ring, we apply a high-voltage to the electrodes for a period of time which can vary between 20 and 70 min. In our current design, the value of the voltage is limited by breakdown between the contact pads, which occurs for 700 V or 2000 V, without and with dielectric oil to increase the permittivity between the probe tips respectively. 

\section{Results}

\subsection{Characterization of optically assisted poling}

\begin{figure*}[t]
    \centering
    \includegraphics[width=\textwidth]{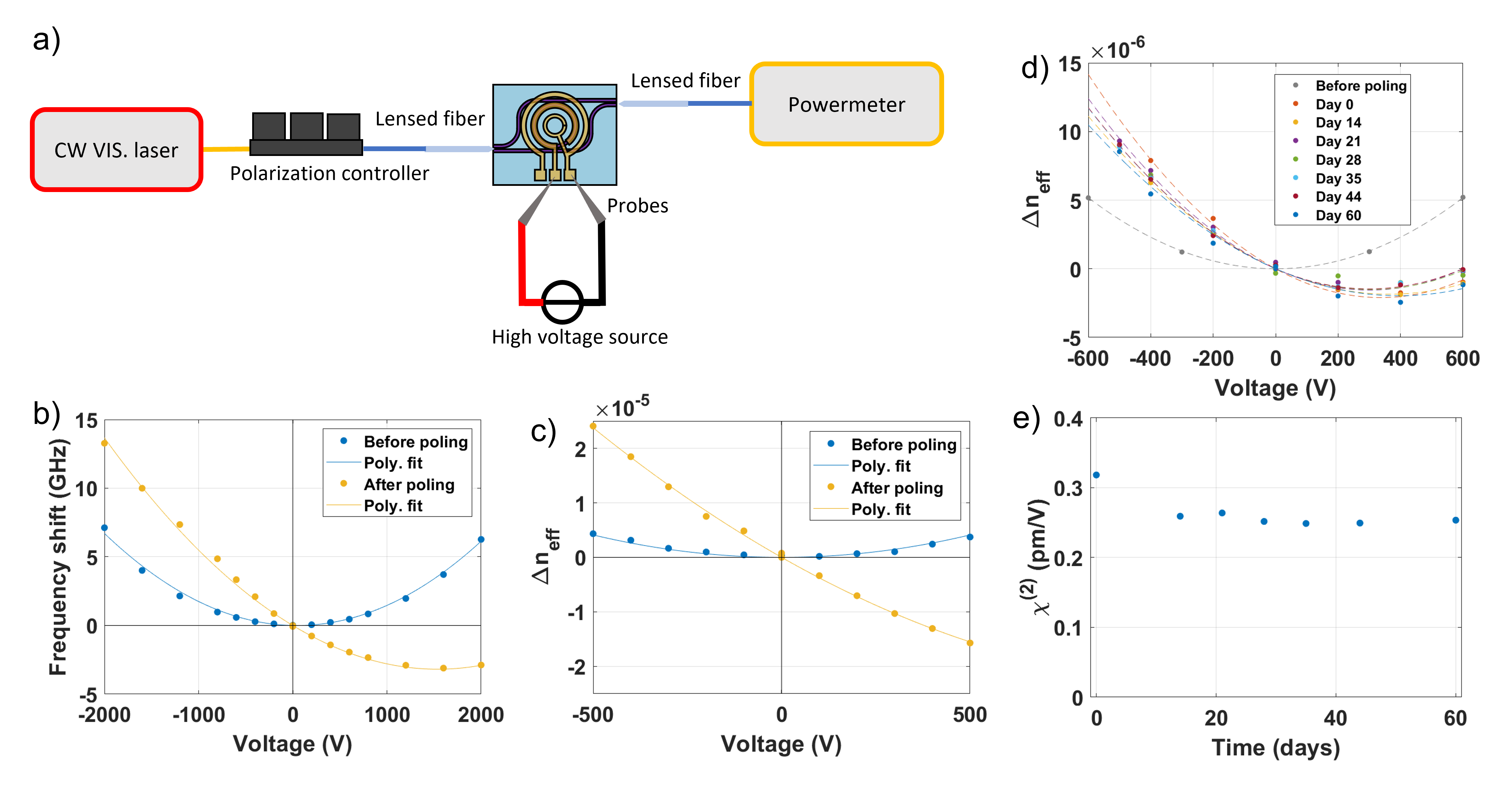}
    \caption{a) Schematics of the poling setup. b) Resonance frequency shift at 1550 nm before and after poling at 2000 V. c) Effective index change at 1550 nm before and after poling estimated from Eq. \ref{eq:delta_f}. d) Resonance frequency shifts measured during 2 months after poling at 600 V at day 0. e) Estimated $\chi^{(2)}_{\rm{eff}}$~during two months after poling at 600 V at day 0.}
    \label{fig:fig2}
\end{figure*}

To characterize the inscribed field, we measured the resonance shift for TM light near 1550 nm as a function of applied voltage, before and after poling as shown in 
Fig.~\ref{fig:fig2} b) 
The data before poling (blue curve) exhibits the expected quadratic behaviour centered at 0 V of the DC Kerr effect. A second order polynomial fit of the experimental data considering Eq.~\ref{eq:delta_f} and \ref{eq:delta_n} gives $\chi^{(3)}$$~\simeq 2.8\cdot 10^3$ pm$^2$/V$^2$. The same experiment was also carried out in TE polarization for which we extract $\chi^{(3)}$$~\simeq 0.96\cdot 10^3$ pm$^2$/V$^2$, which is three times smaller, consistent with Kleinman symmetry rules \cite{Boyd}. From data obtained on five unpoled rings, an average of $\chi^{(3)}$$~\simeq 3.0\cdot 10^3$ pm$^2$/V$^2$ in TM and $\chi^{(3)}$$~\simeq 1.1\cdot 10^3$ pm$^2$/V$^2$ in TE are found. 
After poling the ring with 2000 V applied (average field in the core of 176 V/µm) and an incident 780~nm laser of 250~mW for 30 minutes, the frequency shift behavior (yellow curve) keeps a quadratic shape but shifts relatively to the origin due to the inscription of a permanent DC field indicating the induction of $\chi^{(2)}_{\rm{eff}}$~in the RR. Using the same fitting function as before poling, we obtain $\chi^{(2)}_{\rm{eff}}$~$\simeq 1.2$ pm/V in TM polarization. The experiment was repeated several times with different polarization of the 780~nm excitation and gave the same results, suggesting that the poling process does not depend on the polarization of the excitation light. Fig.~\ref{fig:fig2} c) shows the corresponding effective index variation zoomed in the center region, highlighting the apparition of a nearly linear voltage dependence close to 0 V after poling. It is worth mentioning that this experiment was limited by dielectric breakdown in the oil between the contact pads, but further optimization of electrodes would allow to create a higher electric field in the core. Given the current results and considering the dielectric strength of Si$_3$N$_4$~estimated around 800 V/µm \cite{Sze_1967,Surana_2019}, a maximum value $\chi^{(2)}_{\rm{eff, max}}\simeq 6.5$~pm/V is expected. Such a $\chi^{(2)}_{\rm{eff}}$~lies within the range of nonlinearity of native $\chi^{(2)}$~materials investigated for integrated photonics like aluminum nitride \cite{Liu_Bruch_2021, Gräupner_1992}.

We see that the minimum of the parabola when poled at 2000 V is located around 1700 V, meaning that the final inscribed field is $85\%$ that of the poling field. We investigated how poling voltage influences the final inscribed field and poled a ring in the same conditions but with 600 V. The result is shown in \ref{fig:fig2} d) (red curve). The minimum of the parabola reaches 340 V corresponding to an inscribed field equal to $57\%$ that of the poling field.  We estimate a $\chi^{(2)}_{\rm{eff}}$~of 0.31 pm/V which does not scale directly with the poling voltage. To ensure that the reduced inscribed field is not a consequence of incomplete poling, we carried out the poling experiment for varying poling duration between 20 and 70 minutes. In all cases we extracted the same value of $\chi^{(2)}_{\rm{eff}}$, meaning that the poling already saturates after 20 minutes. This suggests that the charge distribution within the waveguide's core depends on the voltage and charges get more separated at higher voltages. Poling at higher voltages would increase the performance by inducing a field with a higher value, but also with a more homogeneous distribution for increasing the EO overlap.

The long-term stability of the inscribed nonlinearity was assessed by measuring regularly the same poled ring in the same conditions over two months. The frequency shift curves are shown in Fig. \ref{fig:fig2} d), along with the corresponding $\chi^{(2)}_{\rm{eff}}$~ estimated from the polynomial fits in Fig. \ref{fig:fig2} e). No significant decrease of nonlinearity was observed during this period, except for the second point after two weeks. This initial slight decrease can be due to measurement and fitting tolerances, as we can see in Fig. \ref{fig:fig2} d) that the minimum of the parabola (related to the inscribed field) stays roughly at the same position for all the curves.

\subsection{Comparison with thermally-assisted poling}

\begin{figure}[ht]
    \centering
    \includegraphics[width=.5\linewidth]{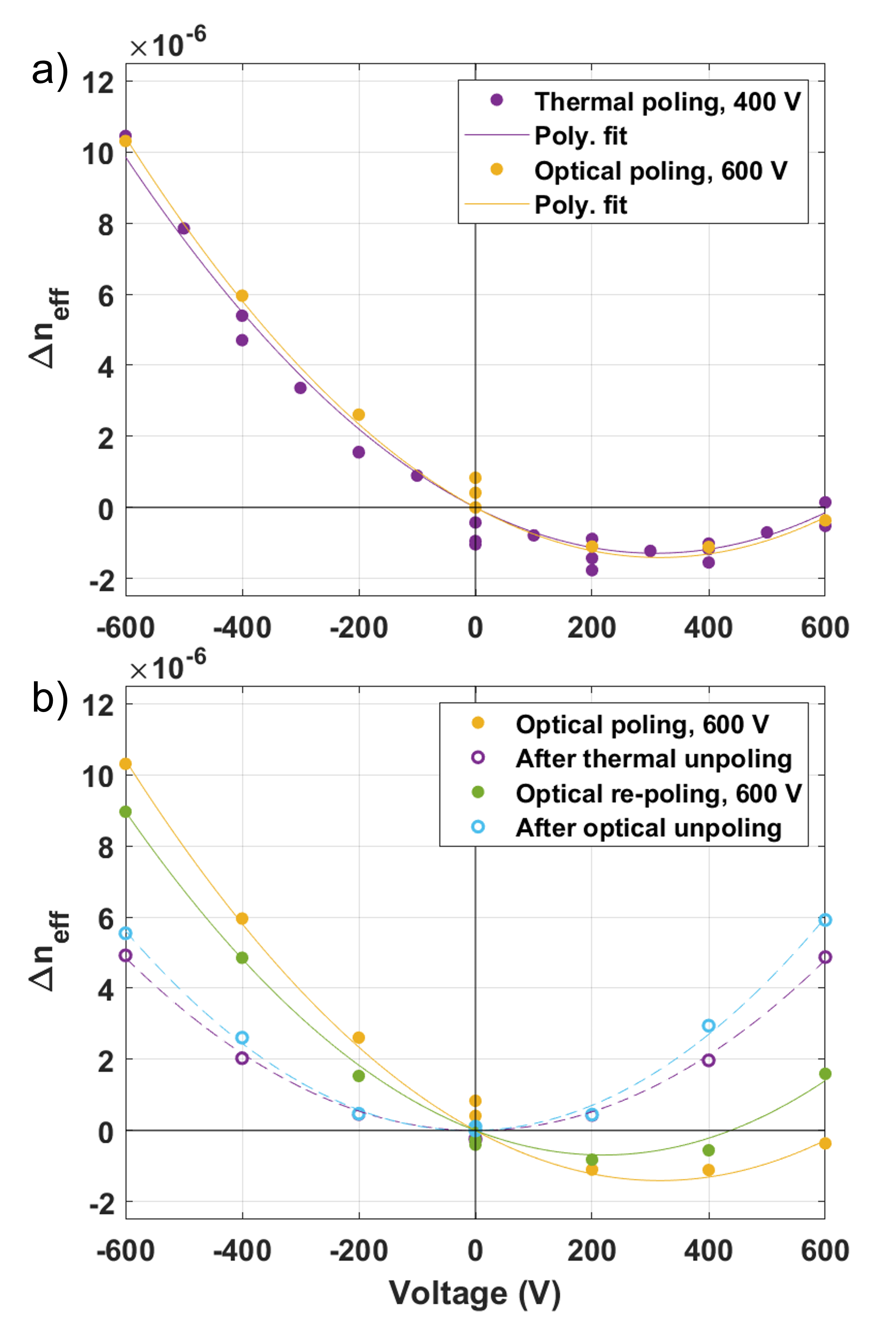}
    \caption{a) Comparison of effective index change between thermally and optically assisted poling. b) Study of inscribed field erasure and re-poling.}
    \label{fig:fig3}
\end{figure}

To compare the efficiency of our optically-assisted method to the previously demonstrated thermally-assisted one, a chip was placed on a hot plate at 260 °C for an hour while applying a high voltage. The nonlinearity was estimated by the measured frequency shift, similarly as in the previous section. The results were then compared to the optically-assisted poled ring as shown in Fig. \ref{fig:fig3} a). We can see that thermal assistance at 400 V gives the same performance as with optical assistance at 600 V. Optically-assisted poling appears slightly less efficient than thermally-assisted poling, but still works within the same order of magnitude of voltage operation. Nevertheless, optically-assisted poling has the advantage of simplicity but also of allowing the application of higher voltages before reaching breakdown between the pads since dielectric strength decreases with increasing temperature, and the application of high permittivity oil is limited at elevated temperatures.

\subsection{Reversibility of the poling process}

An important property to consider is the possibility of erasing and rewriting the inscribed field, bringing more control to the device. Indeed, poled glasses are known to lose their induced nonlinearity when exposed to high temperature or high energy photons \cite{Zabelich_Nitiss_2022, Camara_2015,Pereira_2019}.  
In order to investigate the possibility of thermal unpoling, the previously poled sample was first put on a hot plate at 260 °C for 7 hours and then re-characterized. The effective index change extracted from that measurement is depicted in Fig. \ref{fig:fig3} b) with the purple curve. From the centered quadratic shape after the thermal annealing step, we conclude that the nonlinearity vanished, indicating an erasure of the inscribed field due to the exposure to high temperature. Next, the poling process was repeated in the same conditions as the first step, and a similar second-order nonlinearity was recovered, showing the repeatability of the optically-assisted poling process after erasure (Fig. \ref{fig:fig3} b), green curve). The discrepancy between the originally poled RR and the repoled RR is most likely due to a variation of input coupling of the seed light at 780 nm because of the multimode behaviour at short wavelengths. Finally, the RR was exposed to the same light that was used to pole it for 7 hours. A first check was performed after the first 2 hours and no significant change was observed. After the full 7 hours, the inscribed field was totally erased again, as shown by the centered quadratic curve in blue in Fig. \ref{fig:fig3} b). The erasure then appears to be much longer than the inscription of the field but the reason for this phenomenon is yet to be determined. Nevertheless, these results show that the process is reversible and the poling can be reconfigured either thermally or optically by redistributing the charges evenly across the waveguide's core. 

\subsection{High-speed EO modulation}

\begin{figure}[ht]
    \centering
    \includegraphics[width=.5\linewidth]{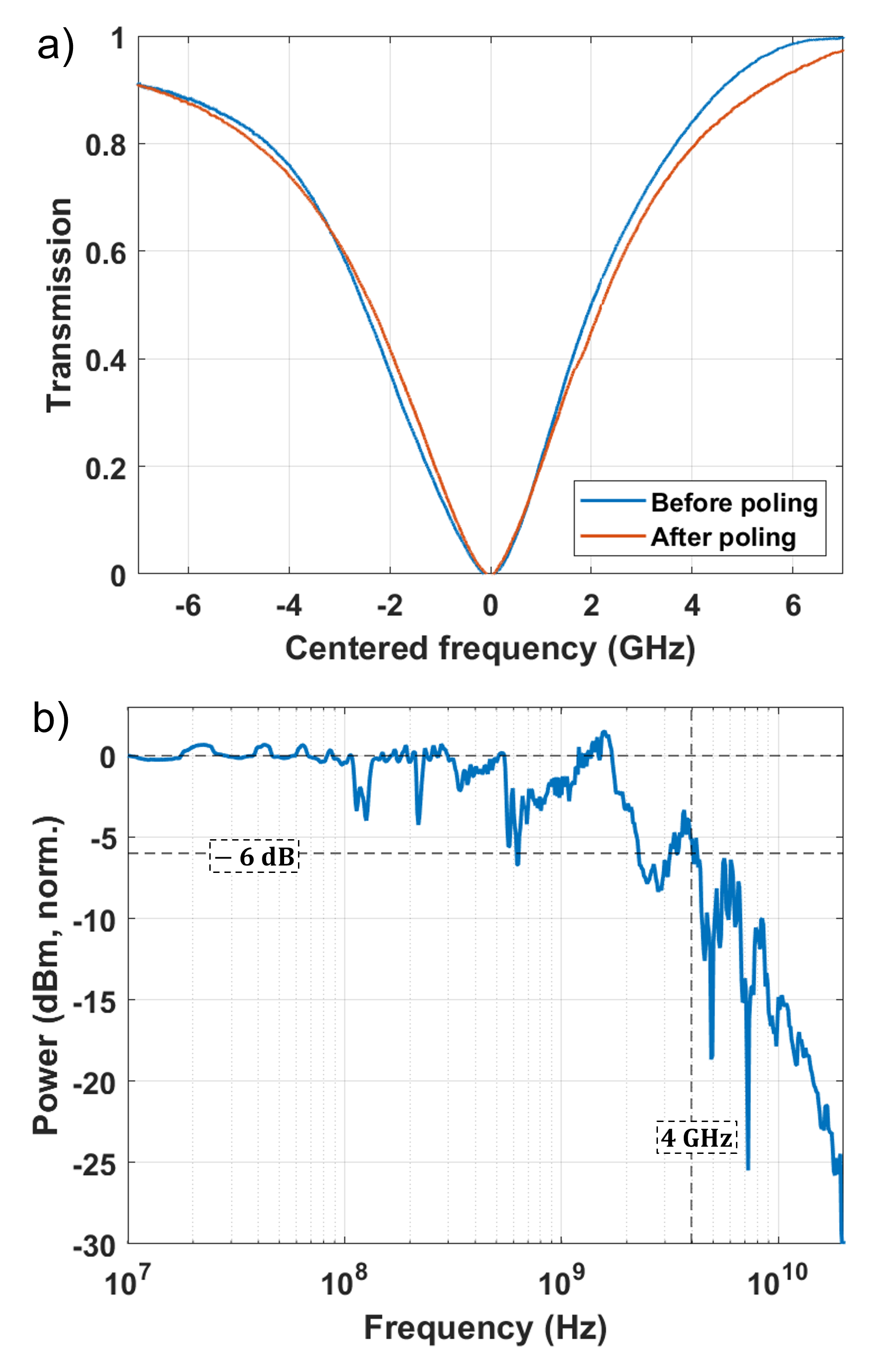}
    \caption{a) Resonance at 1550 nm before and after poling. b) Normalized EO response.}
    \label{fig:fig4}
\end{figure}

High-speed modulation experiments were carried out to show the potential of the poled devices for high-end applications like optical communications. We used the ring with the highest $\chi^{(2)}_{\rm{eff}}$~of 1.2 pm/V and we operated within a resonance at 1550 nm with a Q-factor of $3.9\cdot 10^4$, as shown in Fig. \ref{fig:fig4} a). This ring was chosen to have a Q-factor low enough to allow modulation in the GHz range while providing a modulation depth high enough to perform the experiments without any external amplifiers. The theoretical optically limited EO bandwidth, related to the photon lifetime in the cavity, is given by \cite{Witzens_2018}:

\begin{equation}
    f_c\simeq \dfrac{1+\sqrt{3}}{2\sqrt{3}}\dfrac{f_r}{Q}
    \label{eq:f_c_Q}
\end{equation}
where $f_c$ is the cut-off frequency at which the EO response drops by 3 dB. Given the loaded Q-factor we measured, we predicted a cut-off frequency of $f_c\simeq 4$ GHz for our device. In the experiment, we used a signal generator to send a 23 dBm RF wave to the modulator, with a frequency ranging from 0 to 20 GHz. EO modulation was done at 1550 nm in TM polarization with a power of 20 mW at the input facet. The output optical signal was detected by an amplified fast photoreceiver connected to an electrical spectrum analyzer (ESA). The response of the generator was measured beforehand in order to calibrate the results. Fig. \ref{fig:fig4} b) shows the EO response of the microring modulator up to 20 GHz, with a 6 dB bandwidth of approximately 4 GHz, in good agreement with the expected Q-factor limited cut-off frequency (note that the 6 dB bandwidth is considered here because the ESA gives the electrical power at the output of the photoreceiver, proportional to the square of the actual EO response). Fig. \ref{fig:fig4} a) also indicates that the Q-factor remains unchanged after poling, implying that the EO bandwidth is not impacted by the poling process and it can be accurately estimated prior to the high voltage treatment with common linear characterizations. 

\section{Discussion and conclusion}

In this paper, we report, to the best of our knowledge, the highest $\chi^{(2)}_{\rm{eff}}$~permanently induced in Si$_3$N$_4$~and demonstrate high-speed EO modulation in microring modulators fabricated using common industrial processes. Our findings reveal that it is possible to excite charges optically using visible light, eliminating the need for high-temperature conditions. To gain a deeper understanding of this process and further optimize it, exploring poling at different wavelengths would be beneficial. For example, previous studies have shown that UV light is particularly efficient for silica fibers electric field poling. However, our current study is limited by the RR design, in which the bus-to-ring coupling is strongly dependent on the wavelength. As a preliminary study, we tried the poling experiment with 1550 nm as excitation light instead of 780 nm under an applied voltage of 2000 V. No poling was observed after 1.5 hours with up to 120 mW of optical power in the ring. This hints that the trap energy level is too deep for the charges to be excited by near-IR light. Consequently, it appears necessary to use higher energy light to seed the poling process.
Possible ways of optimization would be to use a Mach-Zehnder modulator (MZM) with Y-junction splitters that are less dependent on the wavelength or to design one port of the add-drop RR to work at the poling wavelength, while the other port is designed for the nominal near-IR operation wavelength. Illumination from the top could also be a way to remove the constraints associated with short-wavelength operation but electrodes must be adapted to ensure they do not shield the Si$_3$N$_4$ waveguide. This method was used in the UV poling of silica fibers, indicating that it is not necessary for the poling light to propagate in the waveguide. This method could be particularly advantageous for modulators operating in TE polarization since in that case electrodes would be placed to the sides of the waveguide to provide a horizontal field collinear with respect to the optical field, while keeping the top surface of the Si$_3$N$_4$~core exposed to short-wavelength light. 
A reliable configuration for short wavelength light excitation would also be beneficial for a more detailed study of the influence of optical power on the poling process as our current design does not permit to know the intensity of visible light coupled to the ring and only gives a qualitative result. Moreover, the performances can still be improved by a careful design of electrodes to achieve a higher electric field in the waveguide before reaching dielectric breakdown. 

To prove the capabilities of the poled modulator, we carried out high-speed modulation experiments up to 4 GHz without using any optical amplifiers. The EO bandwidth can also be extended either by using lower Q-factor RRs or lumped MZM. Balanced MZM would also have the advantage of providing purely linear modulation by canceling the DC Kerr effect in a push-pull configuration.

In summary, we report optically assisted poling of Si$_3$N$_4$~modulators, demonstrating the possibility of achieving a second-order nonlinearity at the pm/V level. The poling process is performed at room temperature on a regular optical test bench with industrial-grade integrated Si$_3$N$_4$~devices. Our experiments resulted in a long-lasting inscribed electric field that was leveraged to perform GHz EO modulation in the C-band telecom window. 
The performances reached with this process are similar to the more complex and more limited thermally assisted poling. This simple method provides a new way of inducing a high second-order nonlinearity in Si$_3$N$_4$, a necessary property to improve the functionalities of the Si$_3$N$_4$~photonic platform.

\section*{Acknowledgment}

The authors thank Dr. Ozan Yakar and Yesim Koyaz for the fruitful discussions.

\section*{Disclosures}

The authors declare no conflict of interest.

\section*{Data availability}

Data underlying the results presented in this paper are not publicly available at this time but may be obtained from the authors upon reasonable request.

\bibliography{sample}

\end{document}